\documentclass[a4paper,11pt]{article}

\pdfoutput=1

\usepackage{jcappub}

\usepackage[english]{babel}
\usepackage{units}
\usepackage{graphicx}
\usepackage{amsmath}
\usepackage{amssymb}
\usepackage{subfigure}
\usepackage{lineno}

\usepackage{tikz}
\usepackage{pgf}
\usepackage{xspace}

\definecolor{dark-gray}{gray}{0.3}

\newcommand{\xmax}{$\mathrm{X}_{\mathrm{max}}$\xspace}
\newcommand{\vxb}{$\vec{v}\times\vec{B}$}
\newcommand{\vxvxb}{$\vec{v}\times(\vec{v}\times\vec{B})$}

\newcommand{\figwidth}{0.7}

\title{The radio emission pattern of air showers as measured with LOFAR -- a tool for the reconstruction of the energy and the shower maximum}
\author[a,1]{A.~Nelles\note{Corresponding author.}}
\author[a]{S.~Buitink}
\author[a]{A.~Corstanje}
\author[a]{J.~E.~Enriquez}
\author[a,b,c]{H.~Falcke}
\author[a,b]{J.R.~H\"orandel}
\author[a]{J.P.~Rachen}
\author[a]{L.~Rossetto}
\author[a]{P.~Schellart}
\author[d]{O.~Scholten}
\author[a]{S.~ter Veen}
\author[a]{S.~Thoudam}
\author[d]{T.N.G.~Trinh}

\affiliation[a]{Department of Astrophysics/IMAPP, Radboud University Nijmegen,\\ P.O. Box 9010, 6500 GL Nijmegen, The Netherlands}
\affiliation[b]{Nikhef, Science Park Amsterdam,\\ 1098 XG Amsterdam, The Netherlands}
\affiliation[c]{Netherlands Institute for Radio Astronomy (ASTRON),\\ Postbus 2, 7990 AA Dwingeloo, The Netherlands} 
\affiliation[d]{KVI-CART, University of Groningen,\\ P.O. Box 72, 9700 AB Groningen, The Netherlands} 

\emailAdd{a.nelles@astro.ru.nl}

\abstract{
The pattern of the radio emission of air showers is finely sampled with the Low-Frequency ARray (LOFAR). A set of 382 measured air showers is used to test a fast, analytic parameterization of the distribution of pulse powers. Using this parameterization we are able to reconstruct the shower axis and give estimators for the energy of the air shower as well as the distance to the shower maximum.
}

\begin{document}
\maketitle
\flushbottom

\section{Introduction}
Measuring the radio emission of air showers has proven to be a suitable tool to extract information about the characteristics of the primary cosmic rays \cite{Buitink2014,Apel2014}. Key to achieving these results was the thorough understanding of the emission mechanisms. Most of the emission is due to the interaction of the shower with the magnetic field of the Earth, which leads to a changing transverse current in the shower \cite{Kahn1966,Ardouin2009}. The emission created by this mechanism leads to an electric field that is pointing in the direction that is defined by the cross-product of the propagation vector of the shower $\vec{v}$ and the direction of the magnetic field $\vec{B}$ \cite{Scholten2008}. In addition to this emission, the overabundance of electrons in the shower  front induces to a changing current along the shower axis \cite{Askaryan1962}. This current leads to an electric field that is radially pointing away from the shower axis \cite{Aab2014}. Adding both contributions and thereby taking into account the non-unity index of refraction in the atmosphere, results in a non-rotationally symmetric pattern of emission on the ground \cite{Werner2012,Coreas,Alvarez-Muniz2011}. 

It was proposed in \cite{LDF} to parameterize the total power of this emission with a combination of two Gaussian functions. Here, we will discuss the application of the proposed parameterization to a large set of data taken with the Low-Frequency ARray (LOFAR) \cite{LOFAR}. Of particular interest are the predicted correlations of the fit parameters with the arrival direction, the energy and the distance to the shower maximum of the air shower.

\section{LOFAR and the cosmic ray dataset}

\begin{figure}
\centering
\includegraphics[width=\figwidth\textwidth]{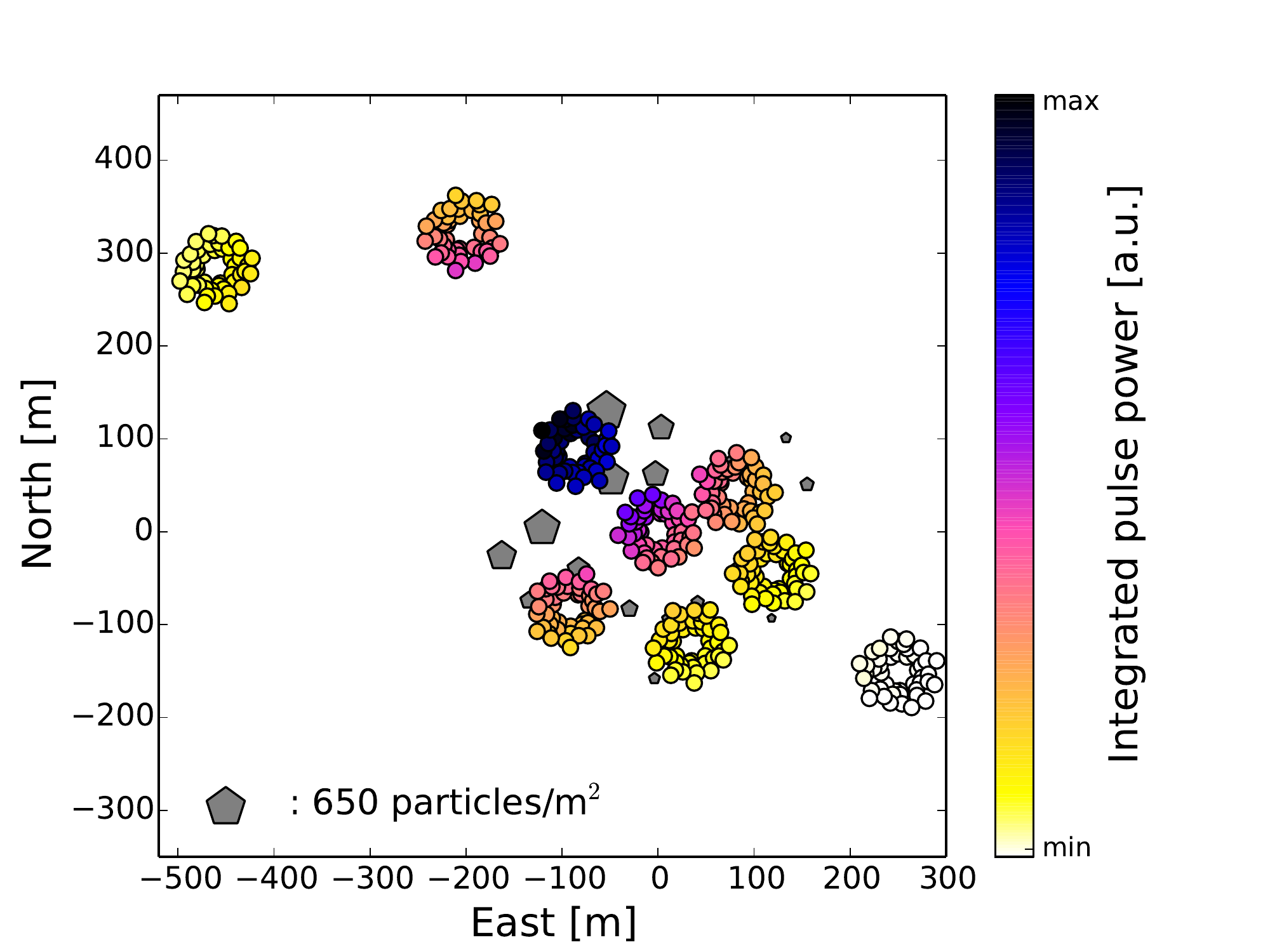}
\caption{An air shower as measured with LOFAR. The shower arrived from an azimuth angle of $162^{\circ}$ (measured northwards from east) and a zenith angle of $37^{\circ}$. The energy is \unit[$2\times10^{18}]{eV}$.  The colored circles indicate the uncalibrated total power as measured with the low-band antennas. The power scales linearly from white to black. The grey pentagons indicate the measured particle densities. The size of the markers scales logarithmically with the detected densities.}
\label{fig:event}
\end{figure}

LOFAR is a digital radio telescope. Its antennas are spread over several European countries and are used together for interferometric radio observations in the frequency range of $\unit[10-240]{MHz}$\ \cite{LOFAR}. The density of antennas increases towards the center of LOFAR, which is located in the Netherlands. Here, about  2400 antennas are clustered on an area of roughly $\unit[10]{km^2}$ with increasing antenna density towards the center. This high density of antennas makes LOFAR the perfect tool to study features of the radio emission created by extensive air showers \cite{LBA}. 

Air shower measurements are conducted based on a trigger received from an array of scintillators (LORA) \cite{LORA}, which results in a read-out of the ring buffers that store the raw voltage traces per antenna for up to $\unit[5]{s}$. LOFAR comprises two types of antennas. While air showers have also been measured in the high-band ($\unit[110-240]{MHz}$) \cite{HBA}, most air showers are measured with the low-band antennas (LBA), which cover the frequency range from $\unit[10-90]{MHz}$. The analysis of the raw data, as well as the characteristics of the air shower measurements are described in detail in \cite{LBA}. The LBAs are not distributed on a grid but in compact clusters of 96 antennas, called stations, as shown in Figure \ref{fig:event}. Of every station either the inner group or the outer ring of antennas (48 antennas each) can be used for cosmic ray measurements at a given moment.

Data recorded with the LOFAR LBAs  between June 2011 and September 2014 are used to test the suitability of the parameterization proposed in \cite{LDF}. The dataset is restricted to those air showers that have been detected in three or more LOFAR stations (clusters of 48 antennas) and that were measured during periods in which no thunderstorm was observed by the KNMI (Royal Dutch Meteorological Institute) in the vicinity of LOFAR \cite{KNMI}. Measured air showers that show a structure in the polarization that is not expected to match cosmic rays \cite{Schellart2014} are also deselected. Showers are excluded if the distribution of the $\vec{v}\times(\vec{v}\times\vec{B})$-component of the unit vectors of the reconstructed electric field shows a large spread ($\sigma>0.23$ per station) or if the average component is larger than $0.45$ in a station. These cut parameters were optimized using the independently identified thunderstorm periods and events containing pulsed radio-interference that did not match the arrival direction as measured with LORA by more than $\unit[30]{^{\circ}}$. All cuts remove 69 events with three or more stations (more than 50\% of these are due to the KNMI criterion) and leave 382 air showers for testing the parameterization. As thunderstorms increase the signal strength \cite{Buitink2007, Schellart2015}, affected showers are more likely to have been detected in more stations, which means that the ratio does not reflect on the duty-cycle of the instrument. 

For every measured air shower the pulse power per polarization and per antenna is calculated by integrating the squared voltage traces in a window of $\unit[55]{ns}$ around the pulse maximum and subtracting the noise contribution. Per antenna, the powers are summed over all three reconstructed polarization directions to obtain the total integrated power. The uncertainties on the power are calculated from the average noise contribution of the $\unit[2.1]{ms}$ of background data that are recorded around the pulse of the air shower.  In addition, a 2\% uncertainty accounting for the antenna model and calibration values is added. This is an estimated uncertainty and only accounts for differences between antennas in a single air shower. Based on the Galactic noise the antennas are already normalized to a common level in the reconstruction process, which reduces the spread between antennas. Furthermore, all signals arrive from (almost) the same direction, in which case an incorrect antenna model will not introduce additional scatter between antennas. Moreover, the analysis is based on the total power, which is less sensitive to the antenna model than the individual components of the signal. When using an absolute scale, differences of about 20\% induced by the antenna model are likely and will have to be considered as additional scale uncertainties. Details of the analysis are discussed in \cite{LBA,wavefront}. The absolute calibration of the data is currently under investigation and will be discussed in detail in a forthcoming publication. 

In order to fit our model to the pulse powers, the antenna positions are transformed into the right-handed coordinate system that is spanned by the direction of the shower $\vec{v}$ and the crossproduct of this vector and the geomagnetic field vector $\vec{B}$, i.e.  $\vec{v}\times\vec{B}$. Here, it has to be taken into account that the reconstructed arrival direction of the air shower is only known with an uncertainty. If a plane-wave is used, this uncertainty is $1^{\circ}$. If the fit of a hyperbola is performed, it decreases to $0.1^{\circ}$ \cite{wavefront}. For this analysis the result of the fit of the wavefront is used whenever possible and uncertainties are propagated accordingly. While the overall direction and its uncertainty are used to determine the coordinate system, individual arrival directions per station are used to reconstruct the signal. 

Since all air showers are measured in coincidence with the particle array LORA, an independent reconstruction of the shower axis and the energy of the shower is available \cite{LORA, LORA_2}. However, as only the most central region of LOFAR is instrumented with these detectors, the delivered accuracy of the reconstruction varies quite significantly, if the shower axis is not contained within the particle array. Still, the position of the shower axis as reconstructed from the particle data is always used as initial value for the fitting procedure of the radio signals. An example of an air shower as measured with LOFAR is shown in Figure \ref{fig:event}.

\begin{figure}
\centering
\includegraphics[width=\figwidth\textwidth]{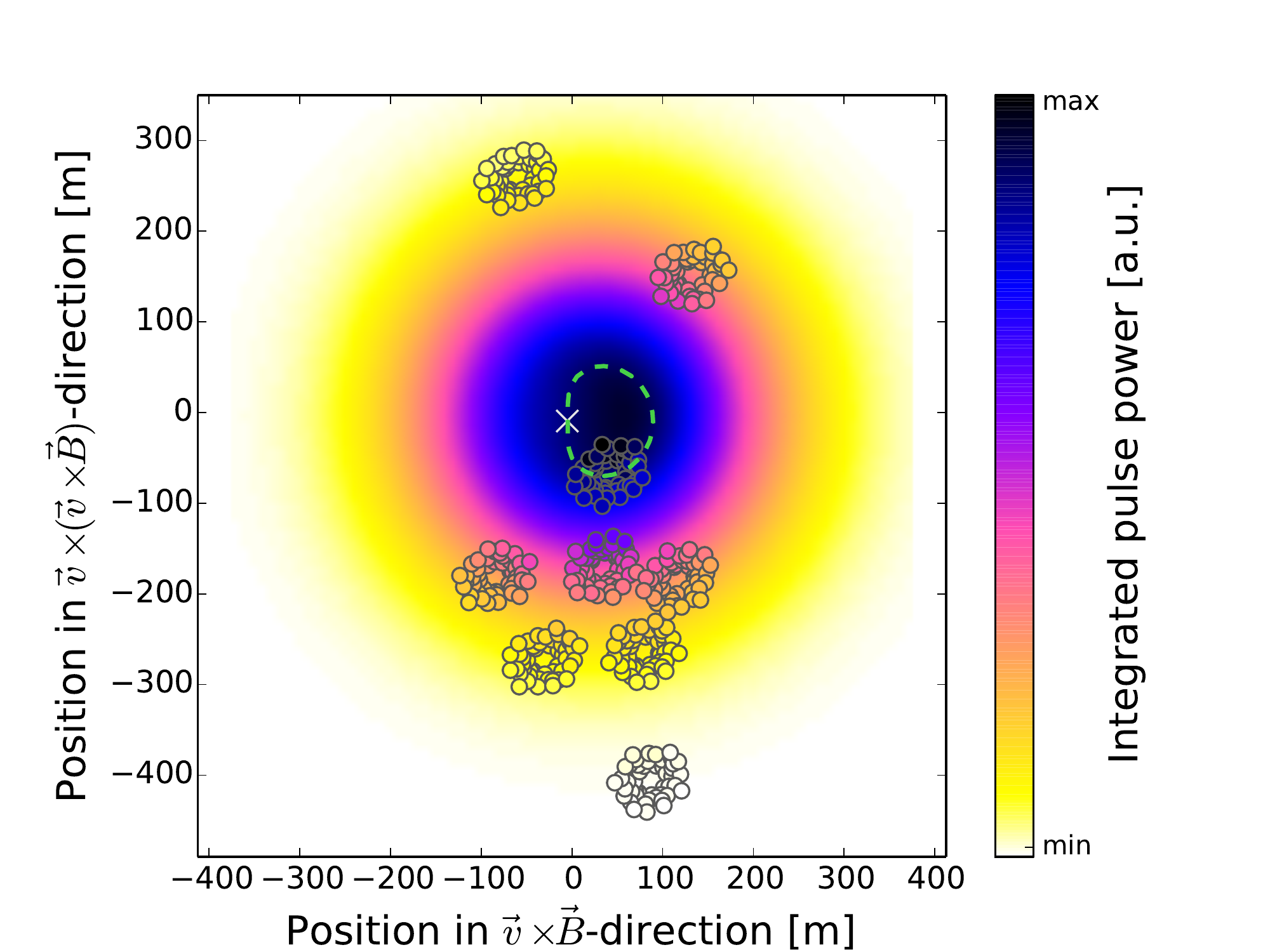}
\caption{The air shower shown in figure \ref{fig:event} transformed into the shower plane with a fit of the signal distribution. The colors in the background represent the fit, while the colors of the circles are the measurements. Also indicated is the shower axis as cross. Both, the axis reconstructed from the particle data and the radio data are at the indicated position. Furthermore, a dashed line of equal signal strength is indicated for the axis position. It emphasizes the asymmetric bean-like shape of the distribution. }
\label{fig:bean}
\end{figure}

\section{Fitting the parameterization to the emission pattern}

For every measured air shower that fulfills the above mentioned criteria the distribution of the total power in the shower plane is fitted with a model. This model has been derived based on the consideration that it has to describe the characteristic \emph{bean-shaped} pattern in the shower plane that is not rotationally symmetric. It has been further refined using CoREAS \cite{Coreas} air shower simulations \cite{LDF}. An example is shown in Figure \ref{fig:bean}.

\subsection{Choice of parameterization and fitting procedure}
In order to fit the selected LOFAR data, the following function is used. 
\begin{eqnarray}
\nonumber P(x^{\prime},y^{\prime})& =& A_+ \cdot \exp\left(\frac{-[(x^{\prime}-X_c)^2+(y^{\prime}-Y_c)^2]}{\sigma_+^2}\right)\\
&& - C_0\cdot A_+\cdot  \exp\left(\frac{-[(x^{\prime}-(X_c+x_-))^2+(y^{\prime}-Y_c)^2]}{(C_3\cdot e^{C_1+C_2\cdot \sigma_+})^2}\right)
\label{eq:fit_lofar}
\end{eqnarray}
The parameters $x^{\prime}$ and $y^{\prime}$ are the coordinates in the shower plane, aligned with the \vxb- and \vxvxb-axis, respectively. Fit parameters are $A_+$, $\sigma_+$, $X_c$, $Y_c$, $C_0$ and $x_-$.

Here, all parameters  that indicate distances and lengths $x^{\prime}$,  $y^{\prime}$, $X_c$, $X_c$, $x_-$ and $\sigma_+$ are given in meters. The parameter $A_+$ is a measure of the energy density in the same units as the measured pulse power. As LOFAR has currently no absolute calibration, the $A_+$ parameter is used for this analysis in instrumental units [a.u.]. The free parameter $C_0$ is dimensionless. The constants $C_1$ and $C_2$ were fixed according to the simulation study of \cite{LDF} to $C_1=2.788$ and $C_2=\unit[0.0086]{m^{-1}}$. The parameter $C_3$ is set to $\unit[1]{m}$.

The fit of equation \ref{eq:fit_lofar} is performed iteratively. First, a single two-dimensional Gaussian ($C_0 = 0$) is fitted to the data to obtain initial values. Consequently, the data are fitted with the complete model using least-squares fitting. It has proven necessary to check whether the initial values deliver a stable result, especially with respect to the choice of $\sigma_+$. Thus, several iterations of this first iteration are performed using different values of $\sigma_+$. The initial values that yield the smallest $\chi^2$ are chosen for the next iteration. 

If there is no set of initial values that produces a non-diverging fit result, the fit is further restricted. The parameter $x_-$ is fixed to a range of $[\unit[-140]{m},\unit[0]{m}]$ and the parameter $C_0$ is fixed to the range of $[0.1,0.7]$. These restrictions allow for a more stable fit result and are slightly larger than the most extreme values that are predicted by the study of simulations. For the fits terminating successfully after the first iteration, $C_0$ is found in the range of $[0.14,0.65]$ with a mean of $0.35$, which is in good agreement with the boundary values and simulations.  The same holds true for $x_-$, which is found in the range of $[\unit[-131.1]{m},\unit[-5.2]{m}]$, showing a weak correlation with the zenith angle, as expected from simulations. More than 50\% of the air showers can be fitted without restriction. This subset shows no bias to preferred arrival directions or fit parameters. The showers are, however, more likely to have been measured with antennas covering wider ranges of azimuthal angles and positions around the shower axis.

In the second iteration of the reconstruction, outliers are removed. Data points that show powers of more than $5\sigma$ from the best fit are rejected. This iteration is necessary to check for antennas with hardware malfunctions. In the whole data-set, never more than 5 antennas are removed per measured air shower. After this cleaning process the data are refitted and the corresponding $\chi^2$ is calculated. 

The uncertainties on the resulting fit parameters are derived from the data. The final fitting step is repeated 300 times on data that are varied within their uncertainties. The resulting parameters are binned and from their distribution the uncertainties can be obtained. Here, the most probable value is identified and the uncertainties are calculated from the 68\% quantile around this most probable value. 

\begin{figure}
\centering
\includegraphics[width=0.49\textwidth]{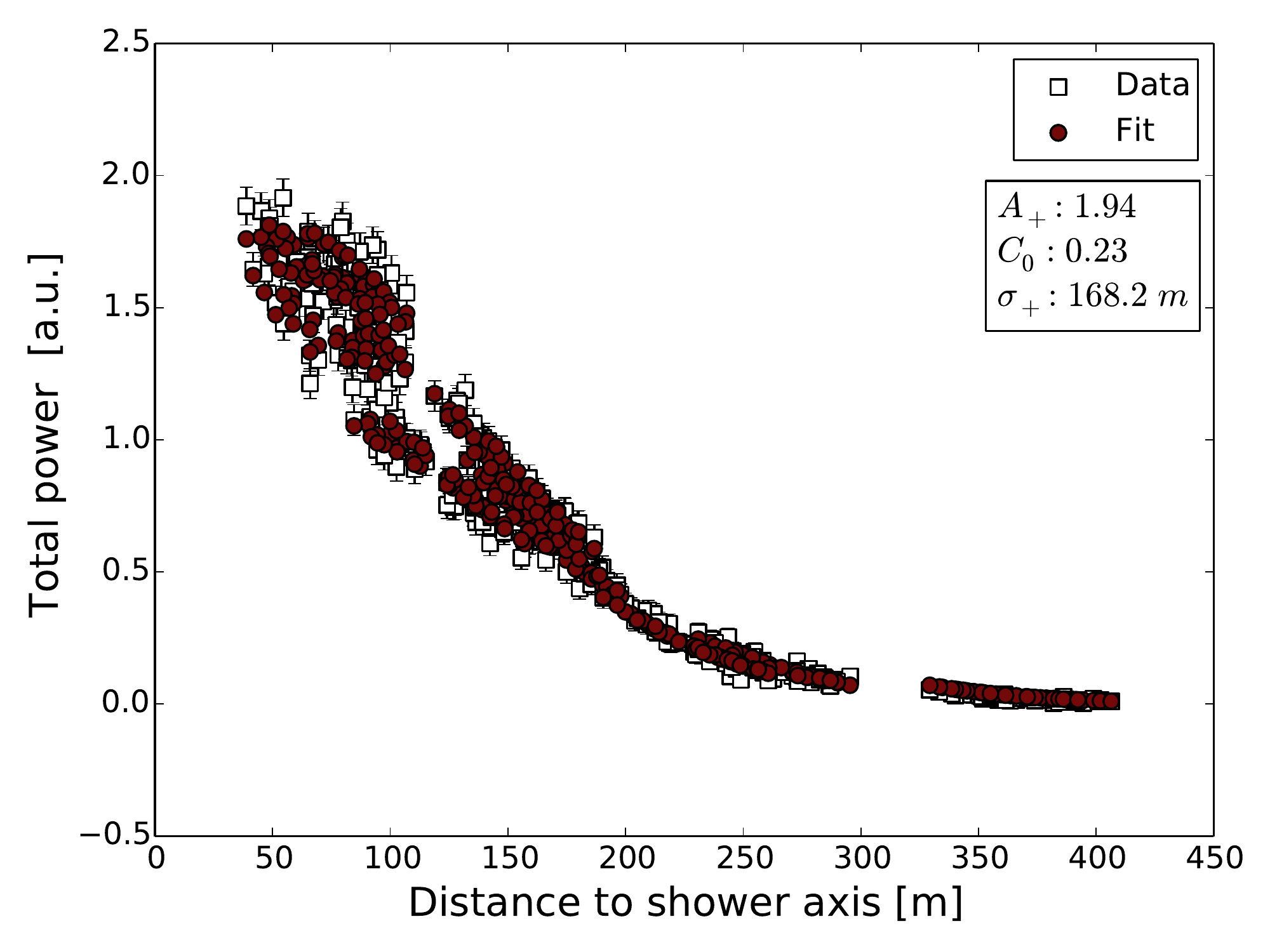}
\includegraphics[width=0.49\textwidth]{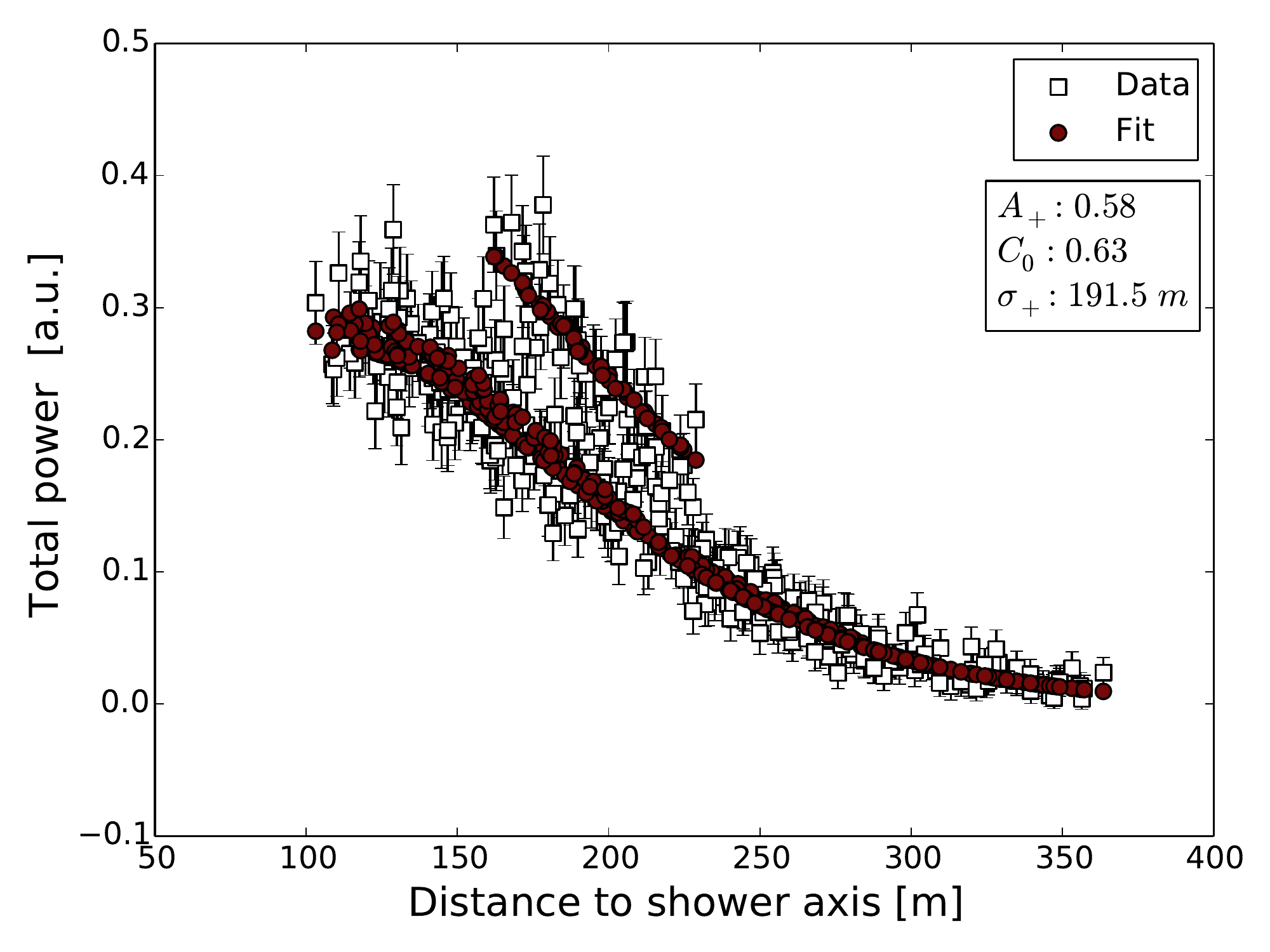}
\includegraphics[width=0.49\textwidth]{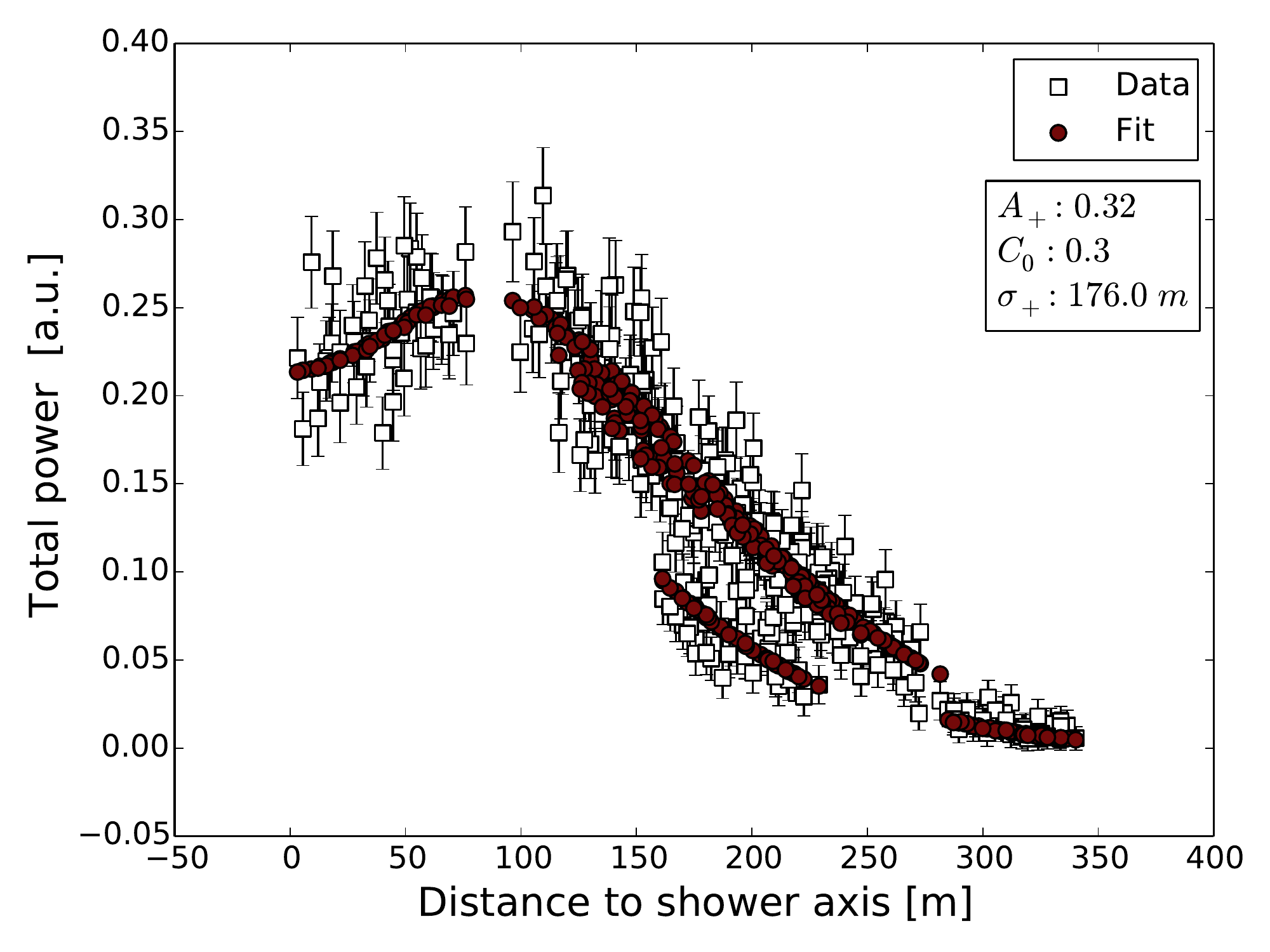}
\includegraphics[width=0.49\textwidth]{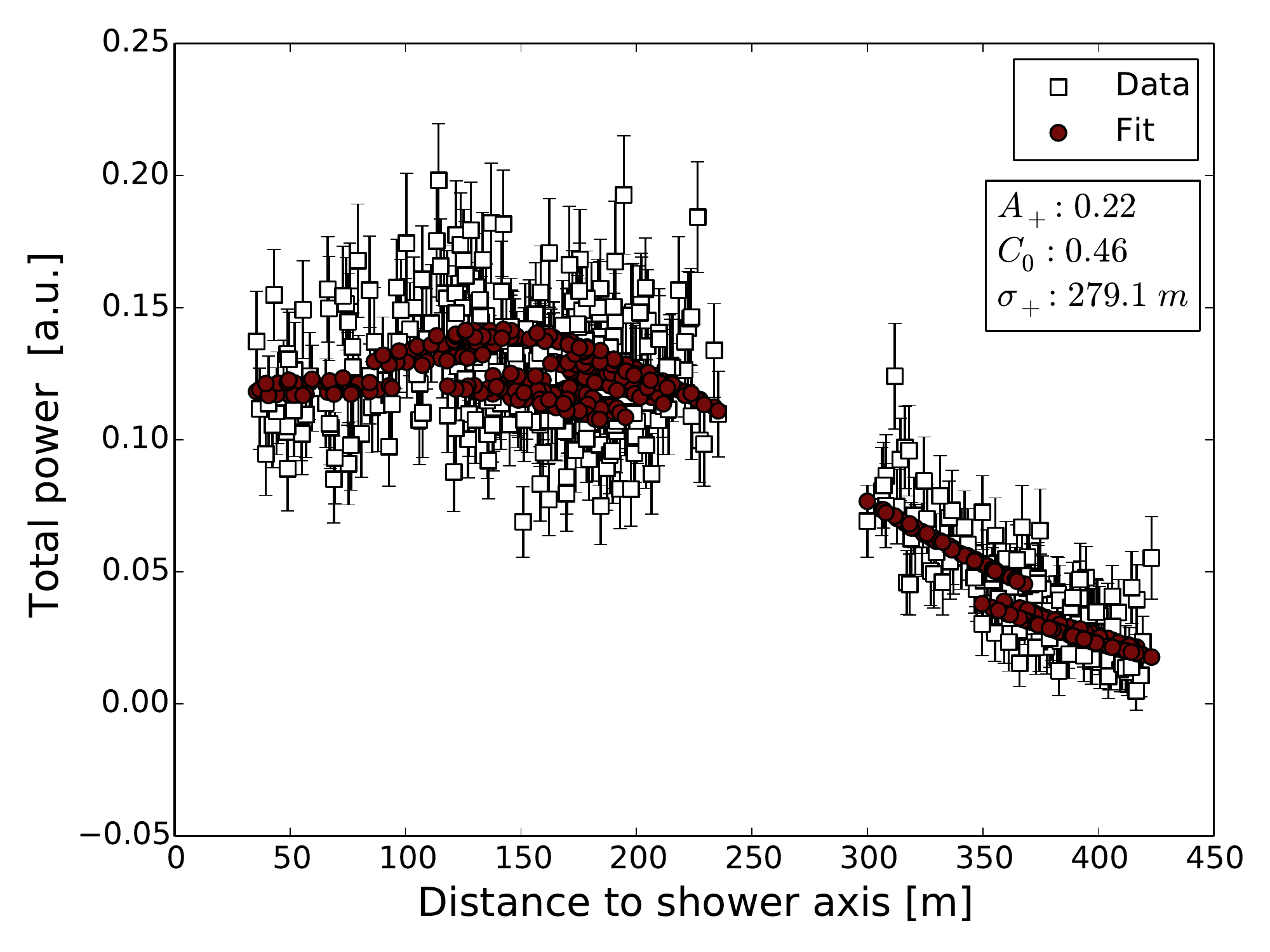}
\includegraphics[width=0.49\textwidth]{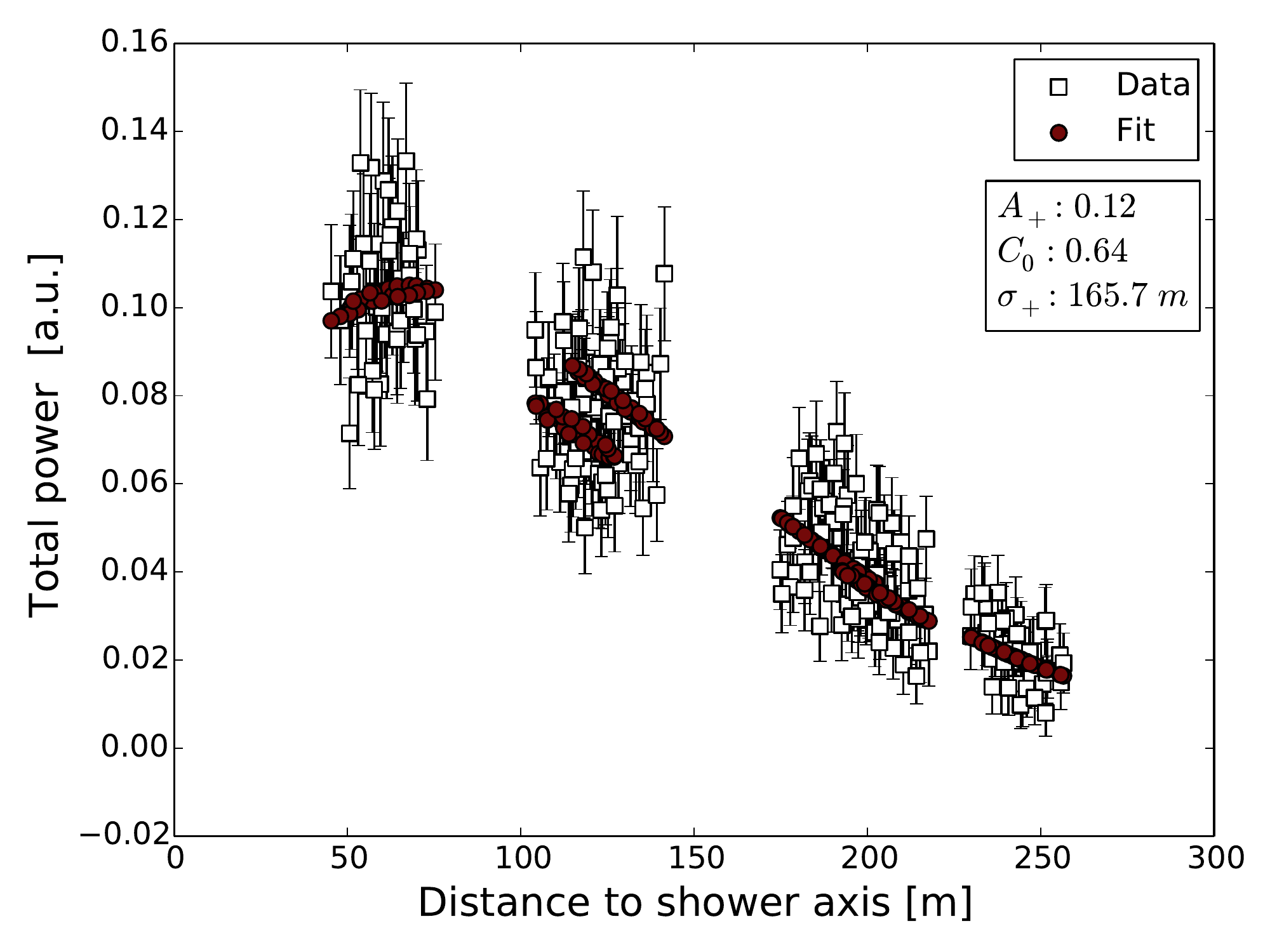}
\includegraphics[width=0.49\textwidth]{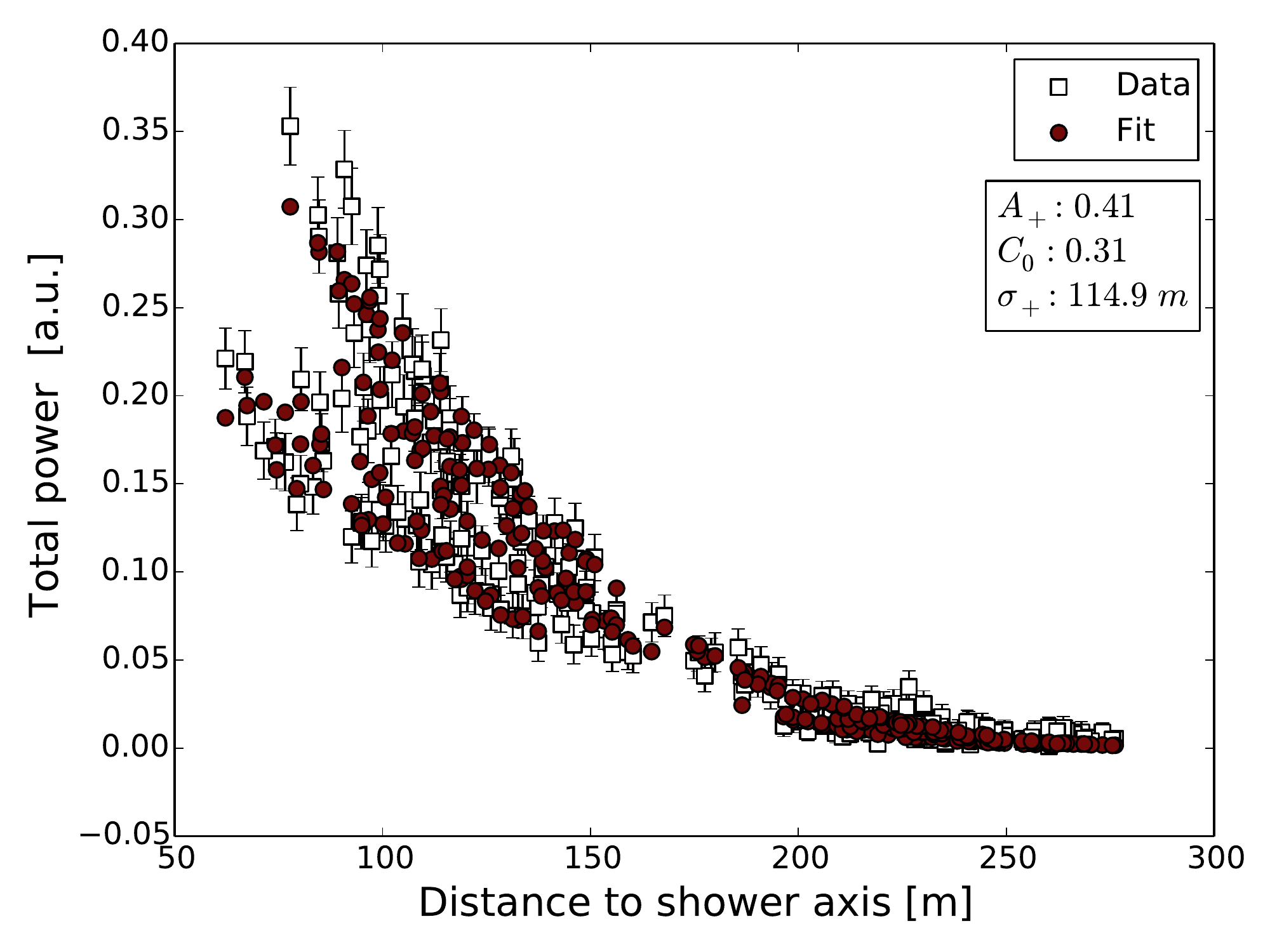}
\caption{Six air showers as measured with LOFAR and reconstructed with equation \ref{eq:fit_lofar}. The total integrated power (measurements: open squares, fit: full circles) is shown as function of the distance to the shower axis. The most important fit parameters are indicated in the box. The asymmetries in the function show that the pulse power is not a simple function of the distance to the shower axis. The figure on the bottom left shows an air shower that was measured with in inner ring of LOFAR antennas, which explains the larger gaps between signal blocks. On the bottom right, an almost vertical air shower ($\theta < 15^{\circ}$) is shown. }
\label{fig:examples}
\end{figure}

\subsection{Discussion of fit quality}

As evidenced in Figure \ref{fig:examples}, the function is capable of reproducing various shapes and distributions of pulse powers that are not compatible with a one-dimensional function of the distance to the shower axis, which assumes a rotational symmetry around the shower axis.

\begin{figure}
\centering
\includegraphics[width=0.49\textwidth]{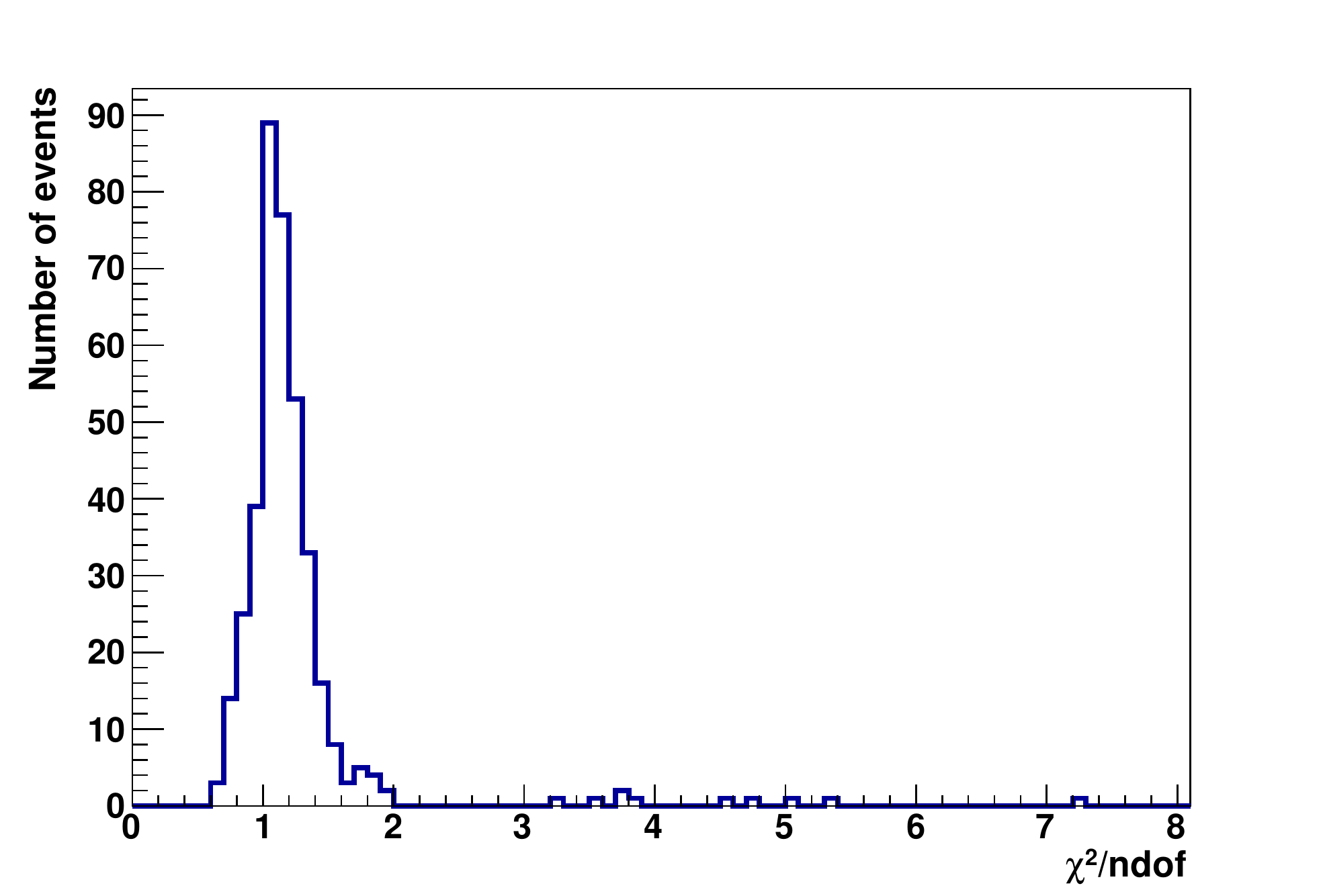}
\includegraphics[width=0.49\textwidth]{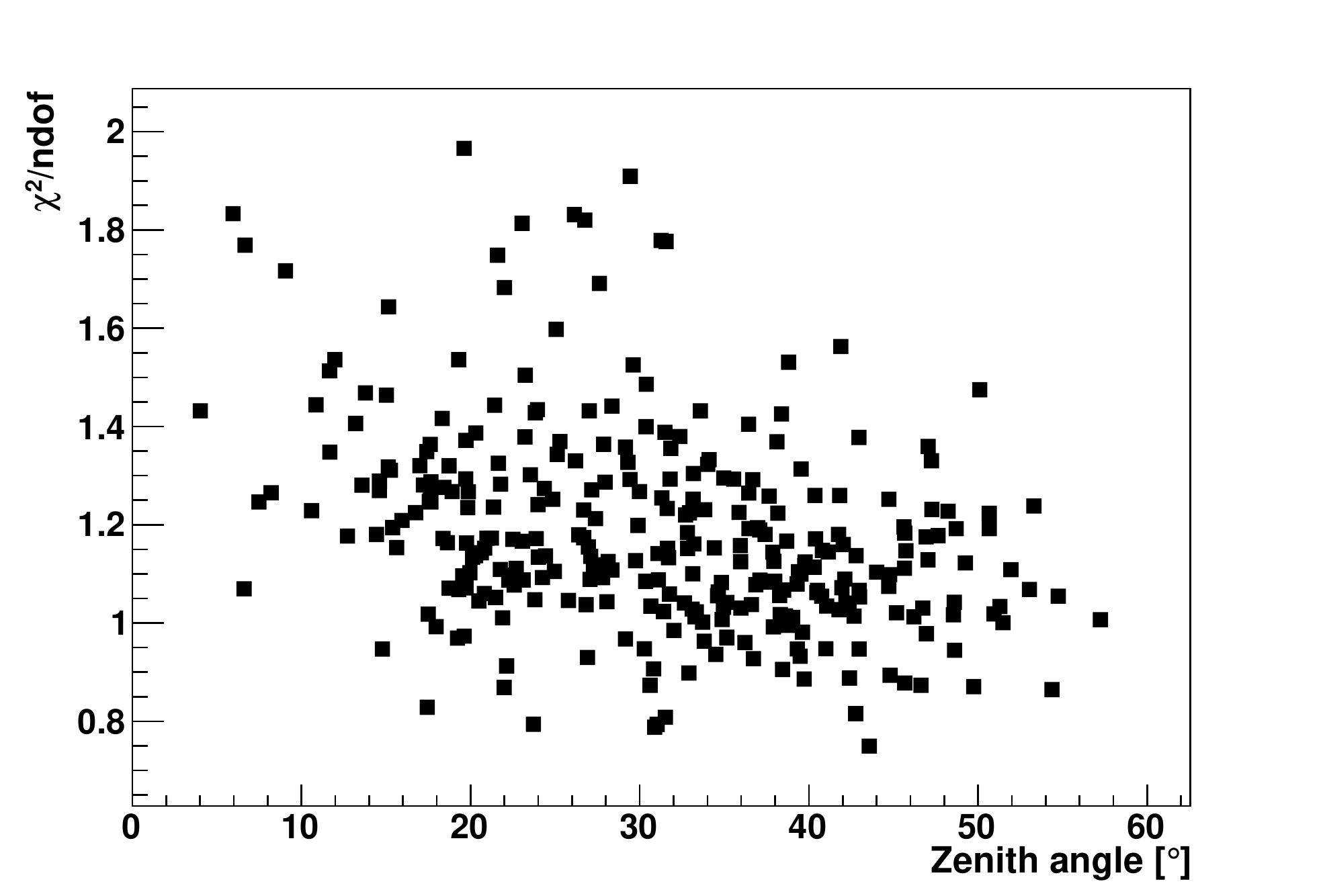}
\caption{Distribution of $\chi^2$ per degree of freedom. Left: Complete distribution of $\chi^2/\mathrm{ndof}$ for the uncut set of fits. The distribution peaks at around one. Right: Values of $\chi^2/\mathrm{ndof}$ as function of the zenith angle of the air shower. The distribution has a correlation coefficient of $-0.3$. }
\label{fig:chi}
\end{figure}

The distribution of $\chi^2/\mathrm{ndof}$-values of all fits exhibits a clear peak at around one, as shown in Figure \ref{fig:chi}.  A number of outliers are visible. Most of these outliers are failed fits, which can also be identified (and removed) through the selection criteria discussed in the following. Furthermore, the function performs slightly worse for showers with smaller zenith angles. This can be seen in the correlation of the $\chi^2$ with the zenith angle, which is depicted on the right in Figure \ref{fig:chi}. An example of such a shower is shown in Figure \ref{fig:examples} on the right in the bottom row. The slope of the fit is smaller than the one observed in data. Thus, the data points closest to the shower axis are reproduced less well, which increases the $\chi^2$ of the fit. The distribution of $\chi^2$-values is rather broad for so many degrees of freedom. This might be explained in two ways. It could indicate a non-perfect fitting model that is less suitable for some types of air showers. However, no correlation of the fit quality with a parameter other than the zenith angle was found. Alternatively, it might be indicative of underestimated event-by-event uncertainties. For example, there has been evidence that the antenna model might not be fully self-consistent for different zenith angles, which could affect the uncertainties differently depending on the background noise and arrival direction. 

A number of cuts have to be introduced to exclude unphysical fits. A large number of problematic air showers are those with a very low signal strength. A very moderate cut of $2.5$ on the averaged ratio of integrated signal and noise power per air shower removes a large fraction of these showers. In addition, two scenarios of unphysical fits are possible: There are two preferred solutions for an air shower or no stable solution can be found. In order to exclude these, the cuts described in table \ref{tab:cuts_ldf} have been implemented. All fits that do not terminate within the given number of 1400 iterations are excluded. In addition, one would like to only consider those showers that have been reconstructed with a reasonable uncertainty, which automatically excludes all reconstructions with two or more stable solutions.  Also, the position of the shower axis should still be in the vicinity of the instrumented area, which can be enforced by a cut on the position of the shower axis. Here, a more sophisticated cut is envisioned for the future. The fit quality decreases for example if there are only antennas on one side of the shower axis, even more so if it is on the side where less asymmetry is visible. At LOFAR, such a cut has to be rather complex given the irregular grid of antennas. For a regular grid such a cut might be easier to implement. Overall, the fit is not successful for 8\% of the selected air showers.

\begin{table}
\centering
\begin{tabular}{lcc}
\hline
\textbf{Cut}&\textbf{Excluded}&\textbf{Air showers}\\
\hline
Air showers in LBA $>= 3$ stations && 382\\
signal-to-noise ratio $> 2.5$&96&\\
$\chi^2/\mathrm{ndof} > 2$&4&\\
$\sigma_{\sigma_+} > \unit[100]{m}$ &11&\\
$C_0 > 1.0$ or $C_0 < 0.0001$ &6\\
$N(\mathrm{function calls})$ at maximum& 5&\\
Remaining for analysis&&291\\
\hline
\end{tabular}
\caption{Quality criteria imposed on the air showers after the reconstruction and on the parameters of the fit. Air showers recorded during thunderstorm conditions have already been removed before. Air showers that  fail any of these criteria are not used for the analysis.}
\label{tab:cuts_ldf}
\end{table}

\section{Reconstruction of shower parameters}
As suggested in \cite{LDF} the parameters $X_c$, $Y_c$, $A_+$ and $\sigma_+$ can be related to physical shower parameters. We assume that the arrival direction of the shower has been obtained through signal timing. The other determining parameters, the position of the shower axis, the energy, and the height of the shower maximum can be determined from the fit parameters. These results are compared to two different sets of benchmark quantities. 
The larger of the two sets are the parameters reconstructed from the signals in the LORA detectors \cite{LORA, LORA_2}. From the LORA detectors a shower axis and an energy is obtained, however, reliably only up to $45^{\circ}$ zenith angle and for a contained shower geometry. The other set consists of the first fifty high-quality air showers detected with LOFAR. For these showers a full Monte-Carlo approach has been chosen, to obtain the position of the shower axis, the energy and the height of the shower maximum \cite{Buitink2014}. In this analysis, the best fitting parameters were obtained by comparing directly and non-parametric CoREAS simulations \cite{Coreas} to LOFAR data. This set of fifty air showers is included in the set that has an independent reconstruction from the LORA detectors. However, due to the limited range in zenith angle, eleven of those fifty air showers cannot be reconstructed reliably with LORA data alone.

\subsection{Reconstruction of the shower axis}
The point of intersection of the shower axis with the ground (position of the shower axis) can be determined from the particle signal, using the fit of an NKG-function \cite{Kamata1958,Greisen1960,LORA}. If the position of shower axis is contained within the area that is instrumented with scintillators (fiducial circle of a diameter of $\unit[150]{m}$ around the centre of the array), an excellent resolution of less than $\unit[10]{m}$ on the position of the shower axis can be obtained \cite{LORA}. However, the scintillators do not cover the complete area instrumented with radio antennas, which results in a decreasing resolution for some air showers. Therefore, the position of the shower axis has to be determined from the radio signal. 

The parameters $X_c$ and $Y_c$ are proxies for the position of the shower axis in the shower plane. However, due to the shift of the maximum signal with respect to the shower axis that is induced by the interplay of geomagnetic contribution and charge excess, the $X_c$ parameter is not identical to position of the shower axis.  It can be corrected using the following equation, derived from simulations \cite{LDF}:
\begin{equation}
x_{axis}~ [\mathrm{m}] = X_c~ [\mathrm{m}]  - (\unit[(28.58\pm0.06)]{m} - \unit[(7.88\pm0.09)]{m}\cdot \sin(\phi)).
\label{eq:core}
\end{equation}

\begin{figure}
\centering
\includegraphics[width=0.49\textwidth]{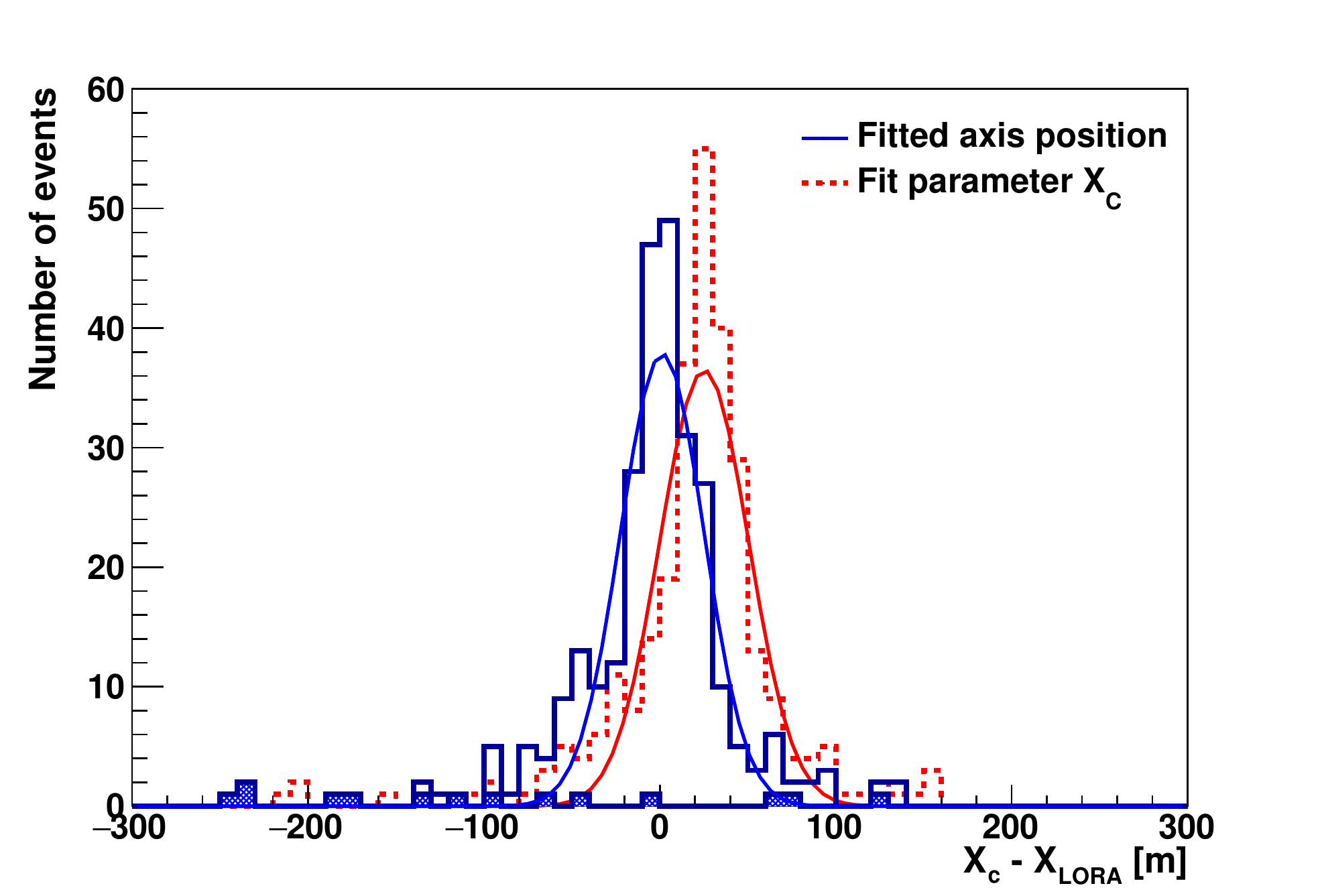}
\includegraphics[width=0.49\textwidth]{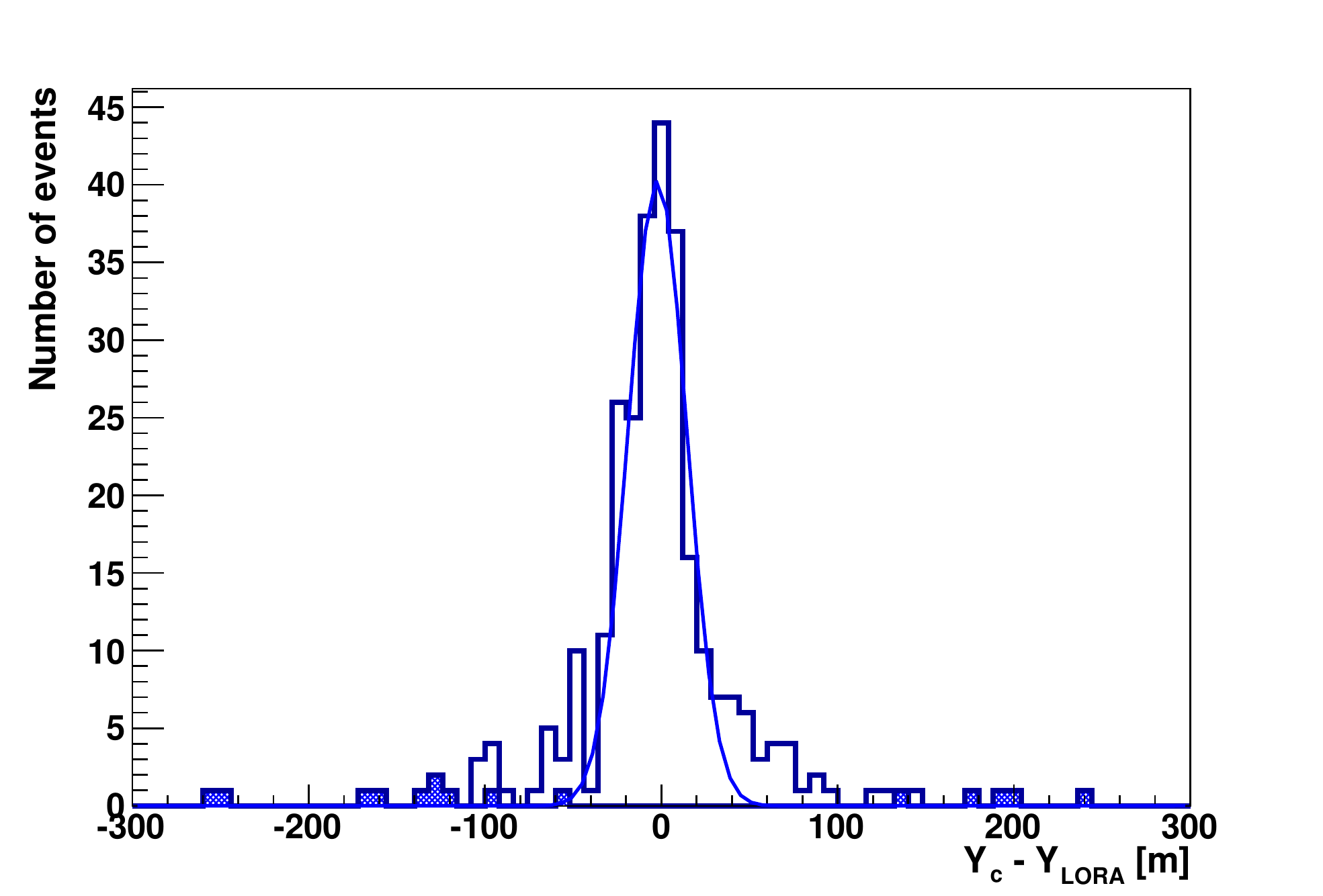}
\caption{Difference between the position of the shower axis as reconstructed from radio data and particle data (LORA). Left: Position in \vxb-direction. The fit parameter $X_c$ shows an offset with respect to the axis position as obtained from the particle data of  $\unit[(25.2\pm1.8)]{m}$. This is indicated by the dotted line and the corresponding fit of a Gaussian. If corrected with equation \ref{eq:core}, the offset reduces to $\unit[(1.3\pm1.9)]{m}$, as indicated by the solid line. The width of the distribution changes from $\unit[(25.3\pm2.8)]{m}$ to $\unit[(23.7\pm2.8)]{m}$.  Right: The $Y_c$ parameter is a direct proxy for the position of the shower axis. The distribution shows a mean of $\unit[(-2.4\pm1.2)]{m}$ and a width of $\unit[(16.5\pm1.1)]{m}$ as indicated by the Gaussian fit. The filled entries of both the histograms indicate axis positions outside of the area instrumented with particle detectors.}
\label{fig:core_lora}
\end{figure}

\begin{figure}
\centering
\includegraphics[width=0.49\textwidth]{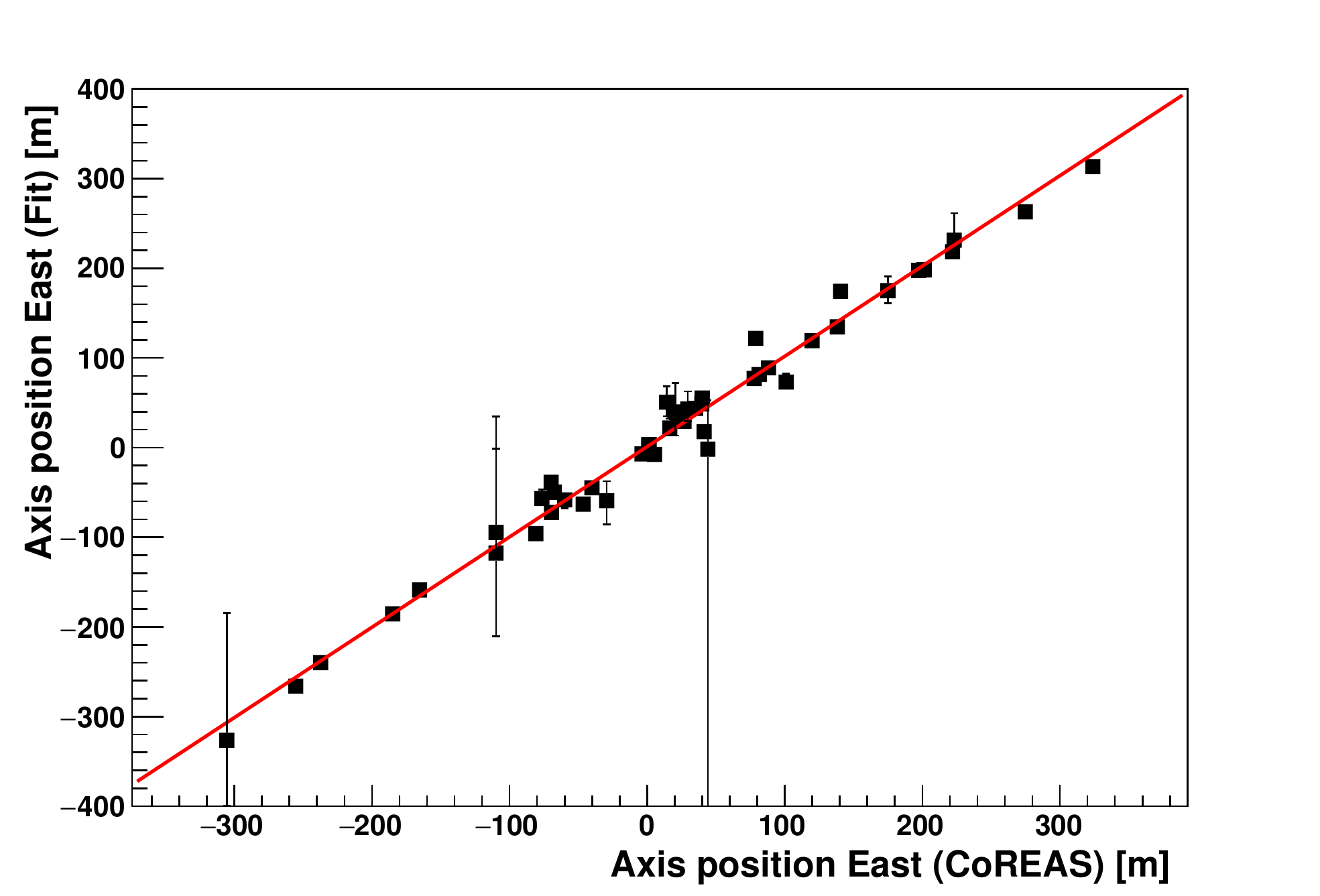}
\includegraphics[width=0.49\textwidth]{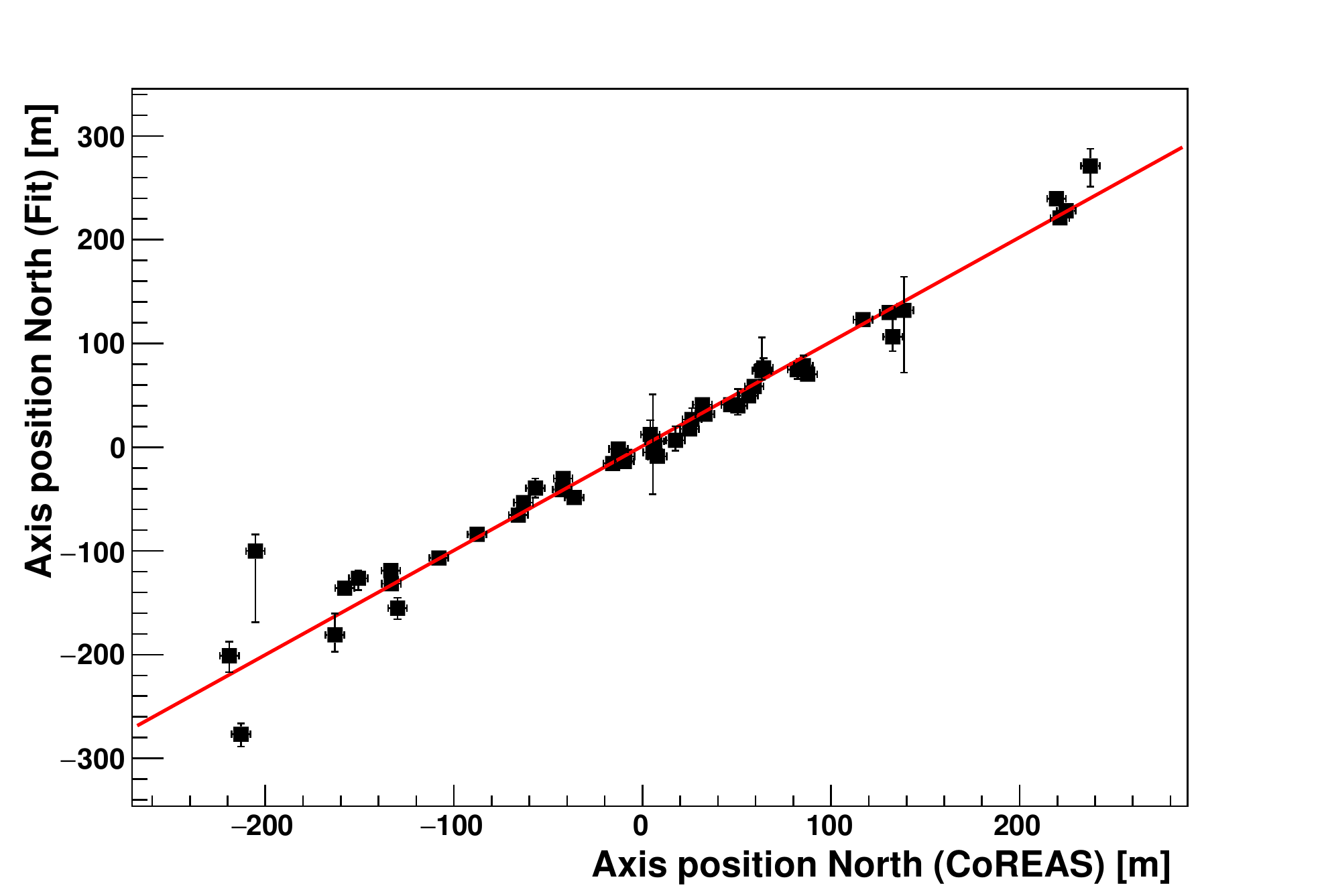}
\caption{The position of the shower axis as obtained in this analysis (Fit) as a function of the axis position obtained from a full Monte-Carlo analysis (CoREAS). For both components of the position (east  in the left and north in the right figure) the best fit of a straight line is shown. The slopes are $0.98\pm0.01$ and $1.01\pm0.01$ with offsets of $\unit[(-0.4\pm0.3)]{m}$ and $\unit[(-1.3\pm0.4)]{m}$, respectively.}
\label{fig:core_coreas}
\end{figure}

The values obtained from data are shown in Figure \ref{fig:core_lora}. In comparison to the LORA reconstruction, the shower axis obtained from the fit to the radio data shows neither a bias nor an unreasonable spread. The resolution in the \vxb-direction ($X_c$) is less good than in the \vxvxb-direction ($Y_c$), which can be attributed to the asymmetry in the signal pattern. If the shower geometry was such, that no antennas were measuring the asymmetry, this shift is difficult to reconstruct and an additional spread occurs. Due to the irregular distribution of antennas at LOFAR, it is impossible to identify a simple cut to exclude this effect. 

Furthermore, it might be argued that the position in \vxvxb-direction shows a small offset towards lower values as predicted in \cite{LDF}. However, the measured difference from zero ($\unit[(-2.4\pm1.2)]{m}$) is not significant. Given the spread on the distribution, it will also be difficult to resolve the effect within these instrumental uncertainties. 

It can also be seen in Figure \ref{fig:core_lora} that the distant outliers are mostly those showers that are reconstructed to have fallen outside the area instrumented with particle detectors. Whenever 20 detectors are active, this area extends to $\unit[150]{m}$ from the centre of the array and showers have an uncertainty of better than $\unit[10]{m}$ on the axis position. The distribution, however, also contains showers that were measured with less than 20 active detectors, which means that the fiducial area reduces accordingly for these showers.

As shown in \cite{Buitink2014} the position of the shower axis is one of the determining factors, when comparing a specific air shower simulation to a measured shower. There, the shower axis, as well as a parameter to correct for the absolute scaling are the only free parameters. Thus, this full Monte-Carlo method provides a high accuracy for the position of the shower axis. Figure \ref{fig:core_coreas} shows a direct comparison between the results from the two-dimensional LDF fit and the direct comparison to CoREAS simulations \cite{Buitink2014}. Covering a range that is much wider than the area instrumented with LORA detectors ([$\unit[-150]{m},\unit[150]{m}]$), both methods agree within their uncertainties. The fitted lines indicate that there is no bias, as the fits are both compatible with a slope of one through the coordinate origin. The spread on this distribution is in both cases about $\unit[12]{m}$. This gives an estimate for the combined resolution of full comparison and this fit for the position of the shower axis.

\subsection{Reconstruction of the shower energy}
The scaling parameter $A_+$ is expected to show a clear correlation with the energy of the primary cosmic ray. This correlation should improve further, if the parameter is corrected for the squared sine of the angle between the arrival direction and the geomagnetic-field, $\sin^2(\alpha)$ to compensate for the dominant emission mechanism. This prediction can again be tested against the two benchmark sets available. 

From the measurements with the LORA detectors an energy can be obtained \cite{LORA_3,LORA_2}. The fitted NKG-function delivers the number of charged particles $N_{ch}$, which can be scaled with the size-energy relation to obtain the energy of the primary particle. In dedicated simulations, the size-energy relations were established for different zenith angle bins. Simulations were performed using both irons nuclei and protons as primary particles, up to $45^{\circ}$ in zenith angle. These relations deliver a band of possible energies, per choice of primary. For protons the LORA detectors deliver a resolution of 40\%, which shrinks to 20\% for pure iron primaries. The largest systematic uncertainty is due to an assumption about the primary composition and thereby the choice of an energy scale. Additional systematics are in the order of 25\%, again differing for an assumption about the composition. 

Details about the energy reconstruction can be found in \cite{LORA_2}. There, a number of selection criteria are suggested that are needed in order to obtain reliable values of the energy of the air shower. The most severe cut is on the position of the shower axis, which needs to be confined within the instrumented area. Furthermore, a cut on the fitted radius of the shower (Moli\`{e}re radius)  is needed.  Imposing those cuts only allows us to retain about 55\% of the original data-set, where the cut on the shower axis is the most severe one. 

However, when fitting the radio signals, also axis positions outside of the area instrumented with particle detectors can be recovered. We therefore chose to use the shower axis as obtained with LOFAR as input for the NKG-fit which allows us to retain more showers at a comparable energy resolution. The cut on the radius, however, still has to be imposed, effectively removing about 18\% of the air showers. Moreover, showers more horizontal than $45^{\circ}$ have to be excluded, based on the missing parameterization and reduced signal strength and increasing fluctuations for more horizontal showers. As air showers are easier to detect by their radio emission at larger zenith angles, the cut at $45^{\circ}$ introduces a reduction of the data-set by roughly 25\%.

\begin{figure}
\centering
\includegraphics[width=0.7\textwidth]{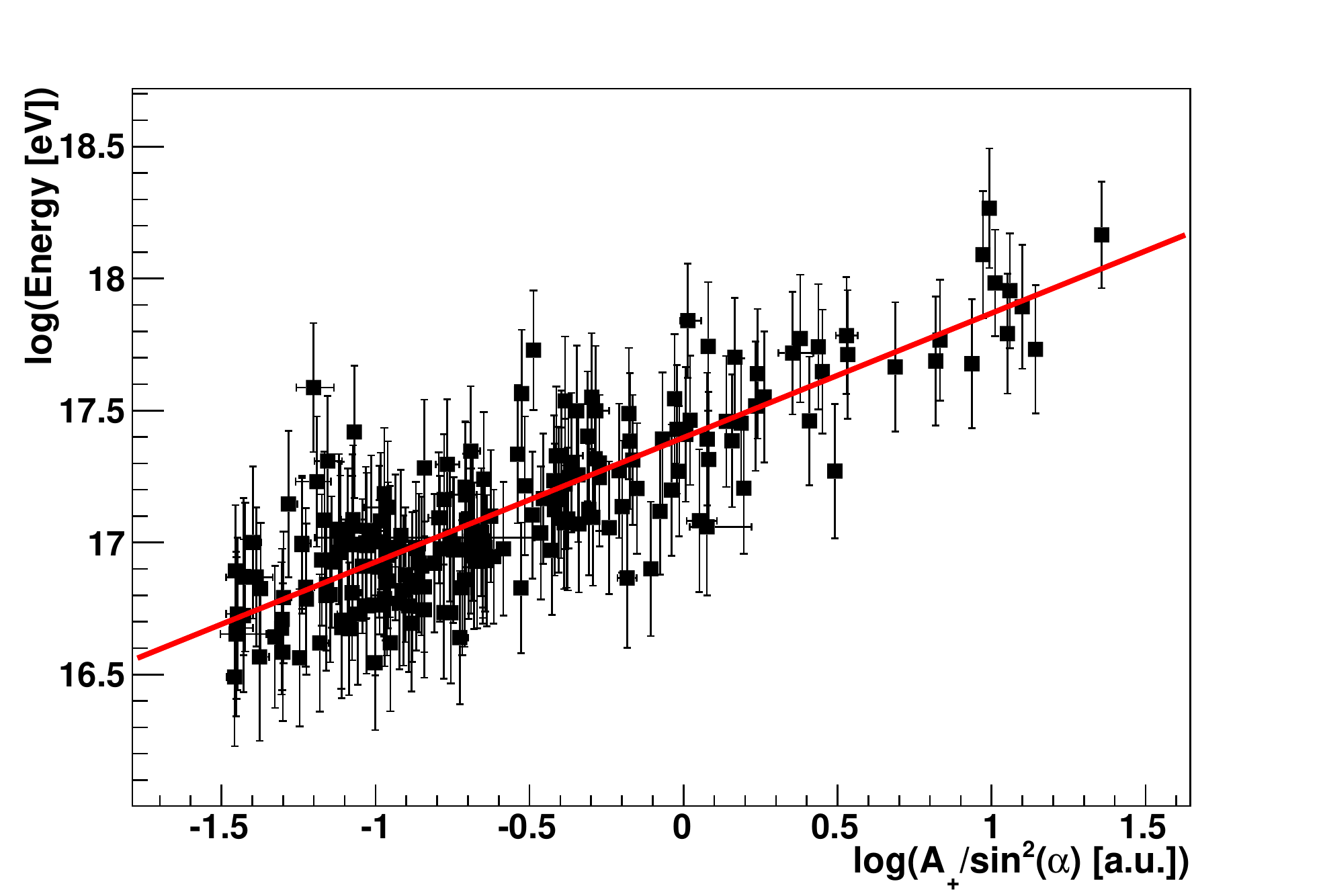}
\caption{Energy as obtained with the LORA detectors as a function of the LDF-parameter $A_+$, which is corrected for the angle between arrival direction and the magnetic field $\alpha$. The zenith angles of the depicted air showers are smaller than $45^{\circ}$. The line indicates the best fit of a straight line with a slope of $0.47\pm0.03$.}
\label{fig:energy2}
\end{figure}

\begin{figure}
\centering
\includegraphics[width=0.7\textwidth]{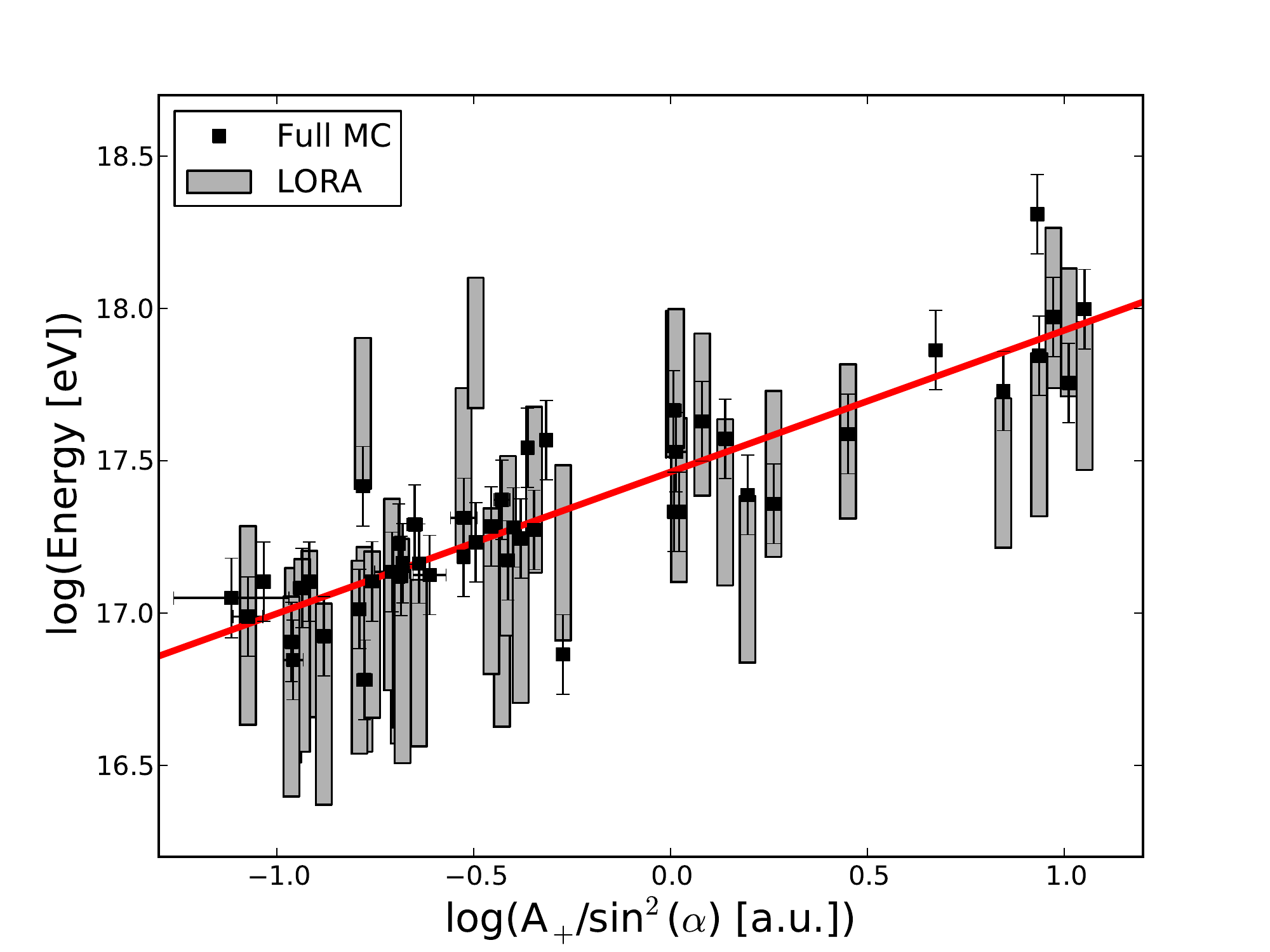}
\caption{Comparison of all energy reconstructions. The energy as obtained from the particle data (grey bands) and the energy from the full Monte-Carlo approach (black squares) are shown as a function of the fit parameter $A_+$. Air showers with no grey bands are more horizontal than $45^{\circ}$. The straight line shows the best fit to the full Monte-Carlo energies. It has a slope of $0.46\pm0.05$.  }
\label{fig:energy1}
\end{figure}

Figure \ref{fig:energy2} shows a clear correlation between the energy estimator based on the radio data and the energy obtained with the particle detectors. The slope of the correlation is about $0.5$, which is expected as the amplitude of the pulse should be proportional to the energy (number of particles) due to coherence \cite{Falcke2005,Huege2008}. Here, the integrated power is used instead of the amplitude, which changes the slope to 0.5. The correlation remains the same within its uncertainties, if reducing the set to those showers that were fit with the unrestricted function, indicating that no bias on the energy estimator is introduced by using a bound fit. The combined energy resolution of the distribution in Figure \ref{fig:energy2} is 41\%. This shows that the resolution is clearly dominated by the energy resolution of LORA.

The fitted energy estimator can also be compared to those energies that have been obtained using a full Monte-Carlo approach \cite{Buitink2014}. Figure \ref{fig:energy1} shows the consistency between all three energy reconstructions. All energies obtained purely from the particle data are consistent with the full Monte-Carlo approach. Furthermore, also here a slope of about 0.5 is found. The energy resolution obtainable from this selected set is 31\%. This is the combined resolution of the full Monte-Carlo approach and of the parameterization. Due to the similarity in approach (both are based on CoREAS), the uncertainties might be correlated. However, the absolute energy scale in the full Monte-Carlo approach is based on the particle signal rather than the radio signal, which excludes a full correlation.  

This illustrates that the energy obtained from the $A_+$-parameter is highly suited as first estimate for the energy of the cosmic ray. The reconstructed energy from the particle detectors is essentially limited by the size of the instrumented area. Furthermore, it is strongly influenced by the unknown primary composition and the lacking sensitivity of particle detectors to this. Using a combined analysis of radio and particle signal will deliver a better handle on the shower-to-shower fluctuations that affect the particle signal. 

\subsection{Reconstruction of the shower maximum}
From simulations \cite{LDF} it is expected that the parameter $\sigma_+$ shows a strong correlation with the distance to the shower maximum, defined as
\begin{equation}
D(X_{\mathrm{max}})\unit{ [g/cm^2]} = X_{\mathrm{atm}}\unit{[g/cm^2]}/\cos(\theta) - X_{\mathrm{max}} \unit{[g/cm^2]},
\label{eq:dxmax}
\end{equation}
where $X_{\mathrm{atm}}$ is the vertical column density of the atmosphere and \xmax is the depth in the atmosphere where the air shower reaches its maximum of secondary particles.

This relation between $\sigma_+$ and \xmax has to be verified experimentally. At LOFAR there is, however, no second detector that can provide an independent measurement of the shower maximum. Consequently, only consistency checks can be performed between different methods.

\begin{figure}
\centering
\includegraphics[width=\figwidth\textwidth]{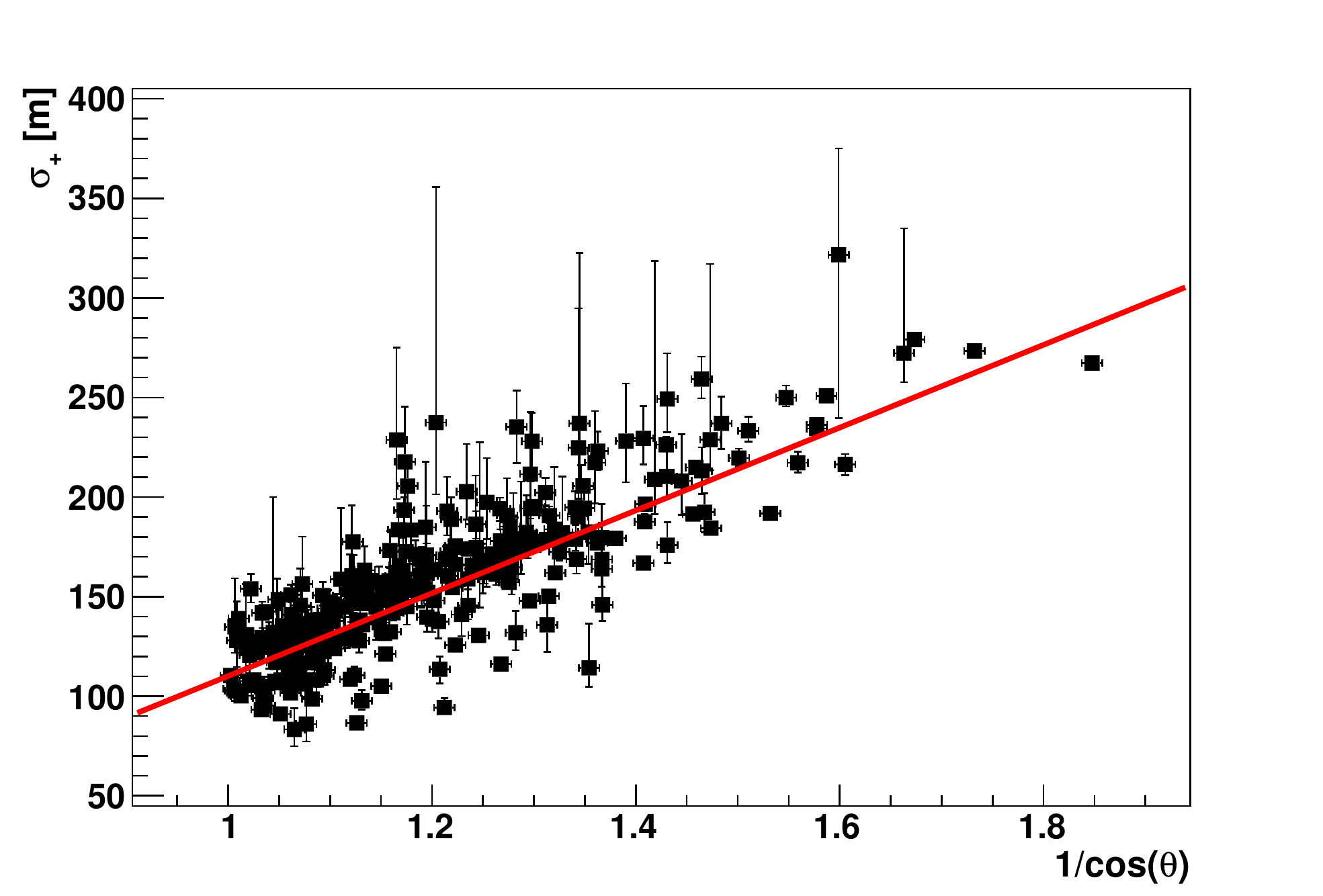}
\caption{All fitted parameters $\sigma_+$ as function of the cosine of the zenith angle $\theta$ of the shower. Also shown is the best fit of a straight line to the data. The spread is expected to be dominated by fluctuations of the shower maximum.}
\label{fig:sigma_zenith}
\end{figure}

The distribution of all reconstructed values of $\sigma_+$ is shown in Figure \ref{fig:sigma_zenith}. The distribution shows that the average value of $\sigma_+$ increases with zenith angle. It roughly follows a $1/\cos(\theta)$ function, which corresponds to what is expected from equation \ref{eq:dxmax}, if $\sigma_+$ describes the distance to the shower maximum. The spread around the fitted line is caused by the fluctuations of \xmax and has the same order of magnitude as expected from the study of simulations \cite{LDF}.

A second consistency check is the comparison to the \xmax-values obtained from full Monte-Carlo approach. In a sense, this test is only a check for self-consistency as the parameterization has been developed based on CoREAS simulations. If those do not describe reality adequately, both results will not be correct. However, given that full air shower simulations reproduce the shape of the signal pattern that has been measured with more than 100 antennas per shower with $\chi^2/\mathrm{ndof}$ values of around one, for two independent simulation codes (CoREAS and ZHAires) a complete mismatch between calculations and measurements is unlikely. 

\begin{figure}
\centering
\includegraphics[width=\figwidth\textwidth]{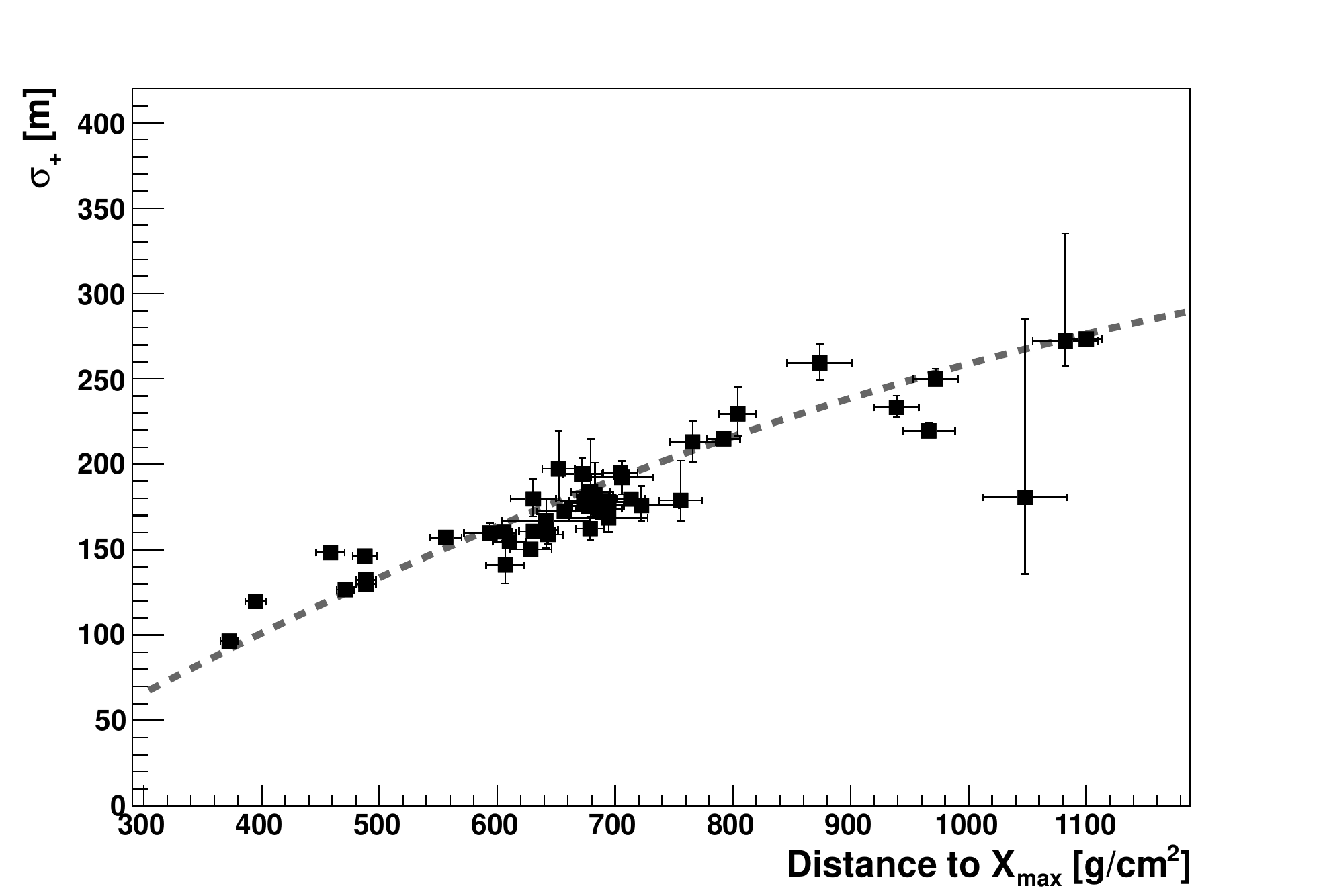}
\caption{Fitted $\sigma_+$ as function of the distance to the shower maximum as reconstructed with the full Monte-Carlo approach for the first 50 air showers measured with LOFAR \cite{Buitink2014}. Also shown in the relation that is expected between $\sigma_+$ and the distance to the shower maximum, as it was predicted from simulations \cite{LDF}.}
\label{fig:sigma_xmax}
\end{figure}

Figure \ref{fig:sigma_xmax} shows the values of $\sigma_+$ as a function of the distance to the shower maximum as obtained from the direct comparison to simulations. The figure shows that both parameters are indeed correlated. The correlation follows almost exactly the relation that was predicted from simulations \cite{LDF}, which is another self-consistency check. Using this curve, the most probable \xmax can be estimated with a resolution of $\unit[38]{g/cm^2}$. This is the combined resolution of both methods. 

\section{Conclusion and Outlook}
We have tested a parameterization that is highly suitable to reproduce the distribution of integrated radio pulse power of air showers measured with LOFAR. The non-rotationally symmetric pattern can be fitted with an excellent quality with 6 free parameters. As the shape is now experimentally confirmed, the number of parameters can be reduced further for experiments with less measurements per air shower.

The parameterization delivers axis position, energy and an estimator for the distance to the shower maximum for every shower. These estimators have been benchmarked against the reconstruction of the particle data and against a full Monte-Carlo approach. The obtainable accuracies on energy and shower maximum are sufficient for studies of large data-sets, however likely not the best precision possible with the radio measurement of air showers, which is better than $\unit[17]{g/cm^2}$ \cite{Buitink2014}.

In the case of LOFAR, the results from the parmeterization can be used as input for full Monte-Carlo studies. By choosing the correct energy and a limited range of \xmax, the number of simulations needed to obtain an \xmax-resolution of better than $\unit[20]{g/cm^2}$ can be reduced by a factor of two to four, which is of significant importance considering the total run times of days per shower. Furthermore, the parameterization is a fast tool for event selection, exposure calculations and performance monitoring.

\section*{Acknowledgements}
The LOFAR cosmic ray key science project very much acknowledges the scientific and technical support from ASTRON. Furthermore, we acknowledge financial support from the Netherlands Research School for Astronomy (NOVA), the Samenwerkingsverband Noord-Nederland (SNN), the Foundation for Fundamental Research on Matter (FOM) and the Netherlands Organization for Scientific Research (NWO), VENI grant 639-041-130. We acknowledge funding from an Advanced Grant of the European Research Council under the European Union's Seventh Framework Program (FP/2007-2013) / ERC Grant Agreement n. 227610.

LOFAR, the Low Frequency Array designed and constructed by ASTRON, has facilities in several countries, that are owned by various parties (each with their own funding sources), and that are collectively operated by the International LOFAR Telescope (ILT) foundation under a joint scientific policy.

\bibliographystyle{JHEP}
\bibliography{BIB_LOFAR}

\end{document}